\documentstyle[aps,eqsecnum,epsf]{revtex}

\def\e{{\bf e}}
\def\g{{\bf g}}

\def\H{{\cal H}}

\def\cL{{\cal L}}

\def\L{{\cal L}}

\def\x{{\bf x}}

\newcommand{\be}{\begin{equation}}
\newcommand{\ee}{\end{equation}}
\newcommand{\bea}{\begin{eqnarray}}
\newcommand{\eea}{\end{eqnarray}}

\newcommand{\Eq}[1]{Eq.~(\ref{#1})}

\begin{document}

\draft

\title{Population Dynamics and Non-Hermitian Localization}
\author{Karin A. Dahmen\cite{addressinfo}, David R. Nelson}
\address{Lyman Laboratory of Physics,
Harvard University, Cambrigde, MA, 02138}
\author{Nadav M. Shnerb}
\address{Racah Institute of Physics, Hebrew University, Jerusalem 91904,
Israel}

\maketitle

\begin{abstract}
We review localization with non-Hermitian time
evolution as applied to simple models of population biology with spatially
varying growth profiles and convection. 
Convection leads to a constant imaginary
vector potential in the Schr\"odinger-like operator which appears in
linearized growth models.   We illustrate the basic ideas by reviewing how
convection affects the evolution of a population influenced by a simple
square well growth profile. Results from discrete lattice  growth
models in both one and two dimensions are presented.    A set of
similarity transformations which lead to exact results for the spectrum and
winding numbers of eigenfunctions for random growth rates
in one dimension is described in
detail.     We discuss the influence of boundary  conditions, and argue
that periodic boundary conditions lead to results which are in fact
typical  of a broad class of growth problems with convection.
\end{abstract}


\section{Introduction}
Bacterial growth in a petri dish, the basic experiment of microbiology,
is a familiar but interesting phenomenon. Depending on the nutrient 
and agar concentration, a variety of intriguing growth patterns 
have been observed\cite{Wakita,Rauprich,Ben-Jacob,Budrene}.
Some regimes can be modeled by diffusion limited aggregation,
others by Eden models and still others exhibit ring structures or a
two-dimensional modulation in the bacterial density. At high 
nutrient concentration 
and low agar density, there is a large regime of simple growth 
of a circular patch (after point inoculation), described 
by a Fisher equation\cite{murray},
and studied experimentally in Ref.\cite{Wakita}.

Of course, most bacteria do not live in petri dishes, but rather in 
inhomogeneous environments characterized by, e.g., spatially varying
growth rates and/or diffusion constants.
Often, as in the soil after a rain storm
(or in a sewage treatment plant), bacterial diffusion
and growth are accompanied by convective drift 
in an aqueous medium through the disorder. By creating artificially
modulated growth environments in petri dishes, one can begin to 
study how bacteria (and other species populations) grow in circumstances
more typical of the real world. More generally, the challenges posed
by combining inhomogeneous biological processes with various types of 
fluid flows\cite{Robinson} seem likely to attract considerable interest 
in the future. The easiest problem to study in the context of bacteria 
is to determine how fixed spatial inhomogeneities and convective flow
affect the simple regime of Fisher equation growth
mentioned above.

A delocalization transition in inhomogeneous biological systems
has recently been proposed, 
focusing on a single species continuous growth model, in which 
the population disperses via diffusion and convection\cite{nadav}:
the Fisher equation\cite{murray} for the population number density 
$c({\mathbf x},t)$, generalized to account for convection and 
an inhomogeneous growth rate, reads \cite{nadav}
\bea
\label{equation-of-motion}
\nonumber
\partial c({\mathbf x},t)/\partial t &=&
D \nabla^2 c({\mathbf x}, t) - {\mathbf v} \cdot \nabla c({\mathbf x},t) \\
& & + U({\mathbf x}) c({\mathbf x}, t)
- b c^2({\mathbf x}, t) \, ,
\eea
where $D$ is the diffusion constant of the system,
${\mathbf v}$ is the spatially homogeneous convection (``wind'') 
velocity, and
$b$ is a phenomenological parameter responsible for the limiting of the 
concentration $c({\mathbf x}, t)$ to some maximum saturation value.
The growth rate $U({\mathbf x})$ is a random function
which describes a 
spatially random nutrient concentration, or, for photosynthetic 
bacteria, an inhomogeneous illumination pattern \cite{nadav}.
If $U({\mathbf x})$ is constant over the entire sample, then the convection
term $- {\bf v} \cdot \nabla c({\bf x}, t)$ can be eliminated by a 
coordinate transformation and has no effect on the growth
of the bacteria \cite{ref-frame}. Only the introduction of a spatial dependence
for the growth rate $U({\mathbf x})$ makes the convection term interesting.

In this brief review, we summarize recent theoretical results
concerning \Eq{equation-of-motion}. In Section \ref{sec:II}, we review
results for a ``square well'' shape for the growth profile 
$U(x)$\cite{oasis}. When linearized about the state of zero population,
\Eq{equation-of-motion} is related to Schroedinger's equation with a
square well potential and analytic results are possible. As we shall
see, large enough convective velocities delocalize bound states
in the equivalent quantum mechanical problem,
and this delocalization is accompanied by escape of real growth
eigenvalues into the complex plane. This important signature of 
delocalization is preserved for {\it random} growth profiles.
A lattice approximation to the continuum \Eq{equation-of-motion},
useful for treating both square well and random growth profiles,
is described in Section \ref{sec:III}. 
Exact results and various similarity transformations for the lattice 
model in one dimension \cite{Shnerb-Nelson-PRL} are summarized in 
Section \ref{sec:IV}. The one dimensional lattice model is used to
explore the effect of 
boundary conditions in Section
\ref{sec:V}. In the appendix we give analytical expressions for the
spectrum of a two dimensional homogeneous system for two different lattices 
and the continuum model.

\section{Results for Square Well Growth Profiles}
\label{sec:II}
\begin{figure}
\epsfxsize=4truein \vbox{ \hskip 1.2truein
\epsffile{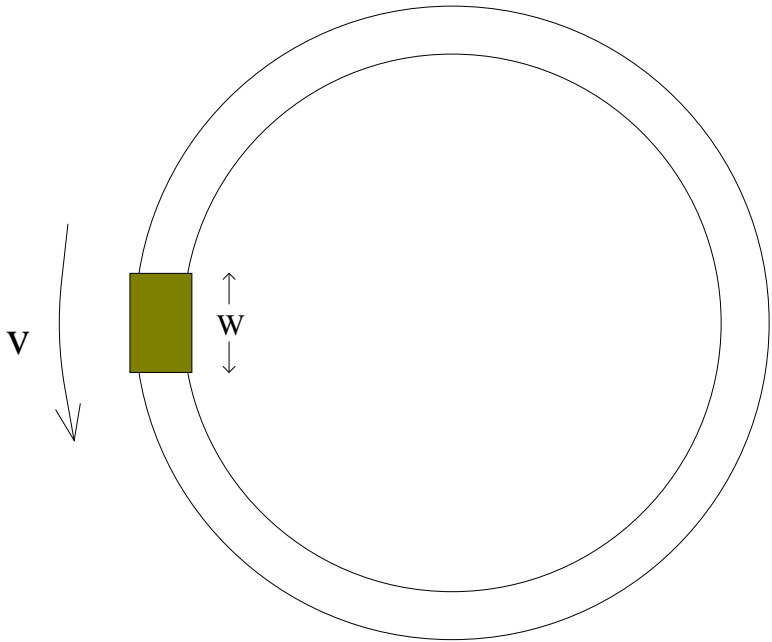}}
\vskip 0.2truein
\caption{Experimental setup:
a solution with photosynthetic bacteria
in a circular pipe or a thin annular track in a petri dish 
is illuminated only in a small area, while
the rest of the sample is either kept dark or illuminated
with reduced intensity. The light source (shaded rectangle) is moved
slowly around the sample to model convective
flow. The bacteria are assumed to divide in 
the illuminated area (``oasis'') at a certain growth rate $a > 0$, 
and die (or grow modestly) in the remaining 
area (``desert'') with growth rate $ - \epsilon a $.
\label{setup}}
\end{figure}

In the following we review results for equation \Eq{equation-of-motion}
with a simple ``square well'' growth profile
$U({\mathbf x})$, imposing a positive
growth rate $a$ on an illuminated patch (``oasis''), 
and a negative growth rate $ - \epsilon a $
outside (``desert'') \cite{oasis}: 
\be
\label{V-def}
%
U({\mathbf x}) = 
\cases{
a\, ,& for $|{\mathbf x}|<{W\over2}\, ,$\cr
\ & \cr
-\epsilon a\, ,& for $|{\mathbf x}|\geq{W\over2}\, ,$\cr
}
\ee
where $W$ is the diameter of the oasis.
Experimentally 
this situation can be realized using  a very simple
setup, which illustrates the basic ideas of localization
and delocalization and leads to interesting further questions.
A one dimensional example is shown in
figure~\ref{setup}, where a solution with photosynthetic bacteria
in a thin circular pipe, or annular petri dish,
is illuminated by a fixed uniform light source through a mask,
leading to a ``square well'' intensity distribution.
The mask is moved at a small, 
constant velocity around the sample to simulate convective
flow. (Moving the mask is equivalent to introducing convective flow 
in the system, up to a change of reference frame\cite{ref-frame}.)
The bacteria are assumed to divide in the brightly 
illuminated area (``oasis'') at a certain rate, but division 
ceases or proceeds at a greatly reduced rate in the darker 
region (``desert'') outside. 
As a result, the growth rate in this continuum population 
dynamics model is positive in the oasis and  
small (positive or negative)
in the surrounding desert region. Using this simple nonlinear
growth model, one can study the 
total number of bacteria expected to survive in the steady state,
the shape of their distribution in space and other quantities,
as a function of the ``convection velocity'' of the light source\cite{oasis}.

Much can already be learned by linearizing
equation~(\ref{equation-of-motion}) around $c({\mathbf x},t)=0$:
\be
\label{linearized-equation}
\partial c({\mathbf x},t)/\partial t = {\cal L} c({\mathbf x},t) \, ,
\ee
with the linearized growth operator 
\be
\label{Liouville}
{\cal L}= D \nabla^2 - {\mathbf v} \cdot \nabla + U({\mathbf x}) \, .
\ee
The convection velocity $\mathbf v$ acts formally like an imaginary
vector potential in this Schr\"odinger-like equation \cite{Hatano}.
For $v$ nonzero, ${\cal L}$ is non-Hermitian,
but it can still be diagonalized by a complete set of right and left
eigenvectors $\{ \phi^R_n({\mathbf x}) \}$ and 
$\{ \phi^L_n({\mathbf x}) \} $,
with eigenvalues $\Gamma_n$,
and orthogonality condition
\be
\int d^dx \phi^L_m({\mathbf x}) \phi^R_n({\mathbf x}) = \delta_{m,n} \, ,
\ee
($d$ is the dimension of the substrate, we focus here on
$d=1$ or $d=2$).
The time evolution of $c({\mathbf x}, t)$ is then given by
\be
\label{time-evolution}
c({\mathbf x}, t) = \sum_n c_n \phi^R_n({\mathbf x}) \exp(\Gamma_n t) \, ,
\ee
where the initial conditions and left eigenfunctions
determine the coefficients
$\{c_n\}$,
\be
c_n=\int d^dx \phi^L_n({\mathbf x}) c({\mathbf x}, t=0) \, .
\ee
(See Refs. \cite{nadav} and \cite{oasis} for a discussion
of how the results from the linearized problem can be used to treat the full
nonlinear system.)

\begin{figure}
\epsfxsize=4truein \vbox{\vskip 0.15truein\hskip 1truein
\epsffile{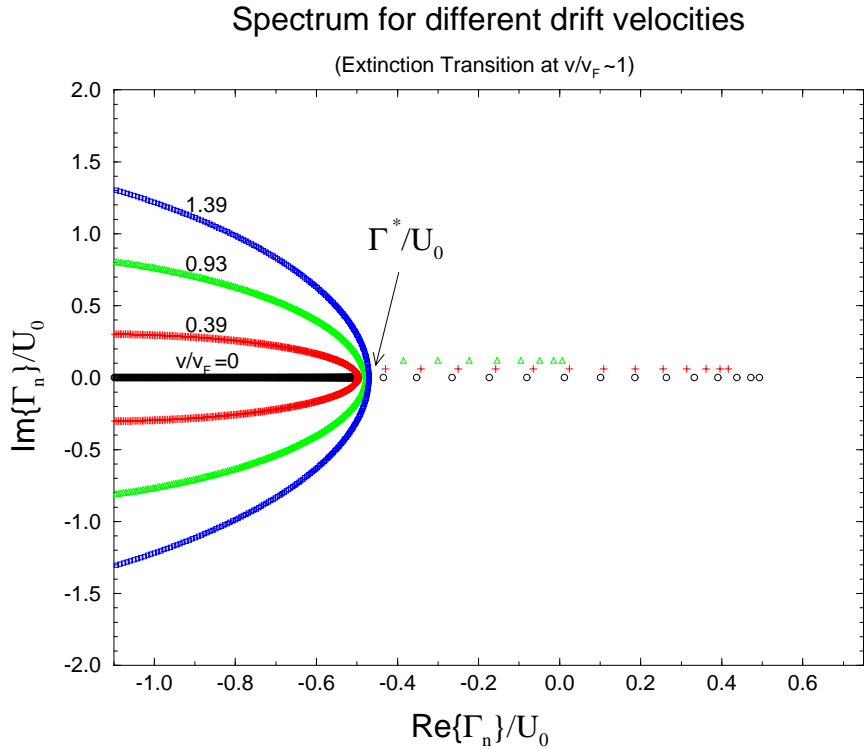}}
\medskip
\caption{Complex nonhermitian eigenvalue spectra 
(normalized by the difference of the growth rates inside
and outside the oasis $U_0\equiv a+\epsilon a = 1$) 
at velocities above and below the extinction transition.
The spectra are extracted from numerical simulations of a 
lattice approximation to the continuum
described in Section \protect\ref{sec:III}, 
for a one dimensional
system of 1000 sites. In units such that the lattice constant 
$\ell_0 =1$, we have
oasis width $W=20$ sites, diffusion constant 
$D=0.3 U_0 \ell_0^2$, with
growth rate $-\epsilon a =-0.5 U_0$ in the desert and $a=+0.5 U_0$ 
in the oasis, 
so that the average growth rate is 
$-0.48 U_0$ (which is equal to the mobility edge
$\Gamma^*$ up to finite size effects).
The velocity parameters 
(in units of the Fisher wave velocity in the oasis 
$v_F= 2\sqrt{aD}$ 
\protect\cite{murray}) 
are $v/v_F\sim 0$ (circles), 
$v/v_F\sim 0.39$ (crosses), $v/v_F\sim 0.93$ (triangles), 
and $v/v_F\sim 1.39$ (squares).
The point spectra are slightly offset in the  vertical direction so as 
to be able to distinguish the eigenvalues of the localized states
for different velocities. As the convection velocity increases,
the mobility edge remains fixed, and the parabola
of the delocalized eigenvalues opens up as $v/v_F$ 
is increased. The real, localized eigenvalues move to the left for
higher velocities. When $v/v_F \geq 1.39$ all states are delocalized.
\label{Gamma-v-series}}
\end{figure}

Fig.~\ref{Gamma-v-series} shows the complex
eigenvalue spectrum associated with \Eq{Liouville} for
a one-dimensional system  with
the potential (\ref{V-def}) and periodic boundary conditions, at
four different values of the convection velocity $v$.
In Ref.~\cite{oasis} details of the derivation of these
results are given.
At zero velocity, $\cal{L}$ is Hermitian and all eigenvalues $\Gamma_n$
are real. There are bound states (discrete spectrum) and 
extended or delocalized states (continuous spectrum).
Because ${\cal L}$ resembles the {\it negative} of a quantum mechanical
Hamiltonian, bound states have the {\it highest} eigenvalues 
(growth rates),
in contrast to the usual situation in quantum mechanics.
At finite velocities, all delocalized states, except the uppermost,
acquire a complex eigenvalue. 
According to \Eq{time-evolution}, 
states with positive real part of the eigenvalue
($Re{\Gamma_n}>0$) grow exponentially with time, states with negative 
real part ($Re{\Gamma_n}<0$) decrease exponentially with time.
In a large one dimensional system 
the ``mobility edge''\cite{mobility-edge}, which 
we define to be the eigenvalue of the fastest growing delocalized state,
(i.e. the rightmost eigenvalue in the 
complex parabolas of figure \ref{Gamma-v-series}),
is located for one dimensional systems of size $L$  
at the overall average growth rate 
\be
\Gamma^* = \langle U \rangle \equiv \int_0^L dx U(x)/L \simeq -\epsilon a \, ,
\ee
In Fig.~\ref{Gamma-v-series},
the eigenvalues of the localized states compose the discrete, real
spectrum to the right of the mobility edge.
With increasing velocity these localized eigenvalues move uniformly
to the left by an amount $v^2/4D$, and 
successively enter the continuous delocalized spectrum
through the mobility
edge, which remains fixed. The parabola also broadens in the 
vertical direction: the imaginary parts of the eigenvalues 
of the delocalized states grow by an amount proportional to $v$.
A given localized right eigenfunction $\phi_n^R$
undergoes a ``delocalization transition''
when the velocity reaches a corresponding critical delocalization
velocity $v=v^*_n$, at which its eigenvalue $\Gamma_n$ has been
shifted so far to the left that it just touches $\Gamma^*$.
At higher velocities it joins the parabola of eigenvalues 
describing a continuum of delocalized states.
The ease with which such a delocalization transition can be observed
experimentally depends on whether there are 
{\it growing} delocalized eigenstates in the system, i.e. whether the
mobility edge has a positive real value or not.
We distinguish three cases, $\langle U \rangle < 0$, 
$\langle U \rangle > 0$,
and  $\langle U \rangle \simeq 0$:

(1) In a large ``deadly'' desert 
($\langle U \rangle \simeq -\epsilon a < 0 $)
all delocalized states die out,
because the mobility edge lies to the left of the origin,
as in Fig.~\ref{Gamma-v-series}.
The growth rate of each localized eigenstate $\phi^R_n$ then
becomes negative at a corresponding ``extinction'' velocity $v_{nc}$
which is smaller than the corresponding delocalization velocity
$v^*_n$. Thus, as convection is increased, the population
dies out before it can delocalize. 
Total extinction occurs 
when the eigenvalue of the localized
``ground state'' (fastest growing eigenfunction of $\cal{L}$)
passes through the origin, {\it i.e.} when
$v=v_{0c}$ (with $v_{0c} > v_{nc}$ for all $n>0$.
The total bacterial population 
of the steady state in the corresponding nonlinear problem goes
to zero linearly with $v-v_{0c}$ \cite{oasis}.

(2) If the average growth rate $\langle U \rangle$ is positive
({\it i.e} for a small enough desert or a small positive growth rate in an
infinite desert),
the mobility edge lies to the right of the origin and the 
delocalization transition 
can indeed be observed at $v=v_0^*$ where the ``ground state'' becomes
delocalized.
One expects to see {\it universal} behavior near
this delocalization transition, since there is a diverging correlation
length in the system, which renders microscopic details irrelevant
for certain quantities. For example, the correlation length $\xi$ of the
spatial density distribution scales as $\xi \sim 1/(v-v_0^*)^\nu$
with a universal exponent $\nu=1$ \cite{oasis}.

\begin{figure}
\epsfxsize=5truein \vbox{\vskip 0.15truein\hskip 0.5truein
\epsffile{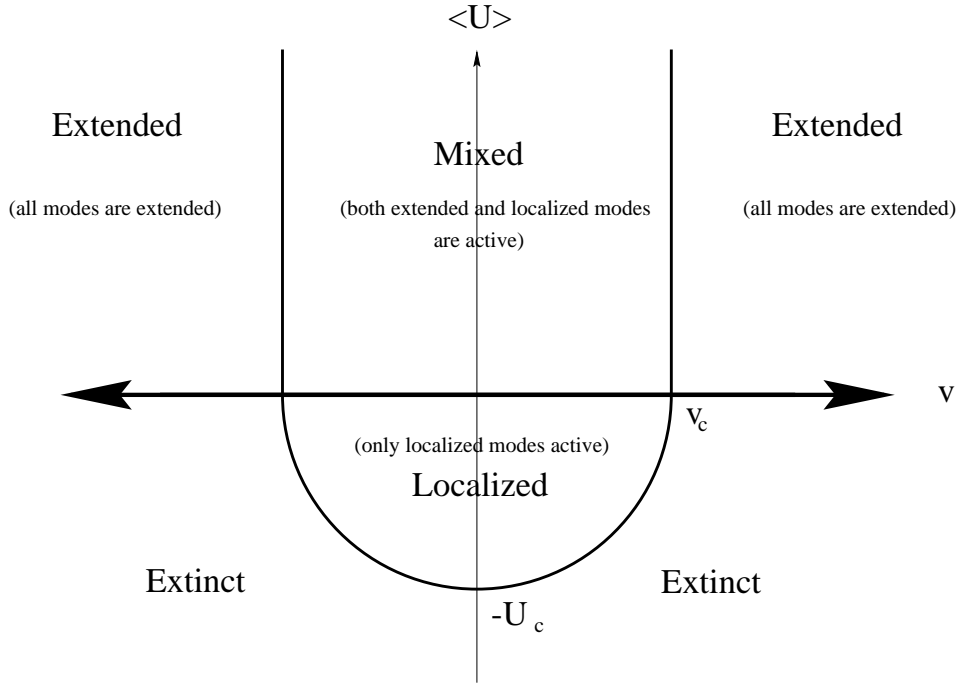}}
\medskip
\caption{Schematic phase diagram in one dimension
for infinite system size, as a function of average growth
rate $\langle U \rangle$ and convection velocity $v$ for 
fixed ``well depth'' $U_0 \equiv a + \epsilon a$.
For a deep well ($U_0 \gg D/W^2$)
the extinction transition out of the localized phase occurs
when $v=0$ for $\langle U \rangle = - U_c$, where $U_c \simeq U_0$.
The diagram shows that if the growth rate is negative outside
and {\it inside} the oasis ({\it i.e.} $\langle U \rangle < -U_c$),
then the only possible state is extinction at any velocity.
If there is positive growth inside the oasis, but negative
outside, a localized population can survive in the oasis,
but only for small enough wind velocities $v$.
Extended states are present for a small positive growth rate
in the desert ($\langle U \rangle >0$). In this case  
localized and delocalized states coexist for small velocities
(``mixed phase''),
while at large velocities all eigenstates are extended.
The ground state becomes delocalized at the critical
velocity $v_c$ which marks the phase boundary
between the mixed and the extended phase.
\label{phasediagram}}
\end{figure}

(3) A special (universal) behavior is expected for the spatially 
average growth rate $\langle U \rangle  = 0$.
In this case the delocalization and extinction velocities coincide.
Fig.~\ref{phasediagram} summarizes the different scenarios
in a sketch of the phase diagram for large systems with fixed well depth 
$U_0 \equiv a+\epsilon a$, obtained by tuning the drift velocity, and 
the average growth rate $ \langle U \rangle \simeq -\epsilon a $.
Also shown in Fig.~\ref{phasediagram} is a horizontal 
transition line at $\langle U \rangle = 0 $ separating a small 
velocity
region ($\epsilon < 0$) where localized modes dominate the steady
state bacterial population, from one ($\epsilon >0$)
containing a mixture of localized and extended states.
It is of course also possible to drive a population {\it extinct}
at zero velocity simply by lowering the average growth rate. This 
special transition at $\langle U \rangle = -U_c$ is indicated at the 
bottom of figure \ref{phasediagram}.

\section{Lattice Model of Growth and Convection}
\label{sec:III}

As discussed in Refs. \cite{nadav} and \cite{oasis}, analytic expressions
for eigenfunctions and eigenvalues obtained from an analysis of a linearized
growth problem can be used to study the nonlinear steady state provided 
only a few unstable growing modes are present.
In more general situations, however, a numerical analysis of the full
nonlinear \Eq{equation-of-motion} may be necessary. Numerical work
also provides insights into situations where the growth profile is random,
as opposed to the simple square well considered in Section \ref{sec:II}.
We review here a lattice approximation to the continuum nonlinear model
which is very useful for obtaining numerical results. It was used
in Ref. \cite{oasis} to obtain growth spectra like those shown
in figure \ref{Gamma-v-series} to confirm and extend analytic results 
obtained in one dimension. Here, we apply the lattice model to some 
illustrative {\it two} dimensional growth problems with convection. 

A discrete lattice approximation (discrete
in space but {\it continuous} in time) to \Eq{equation-of-motion} 
in $d$ dimensions reads \cite{nadav}
\begin{eqnarray}
{dc_{\mathbf x}(t) \over dt}& =& w
\sum_{{\mathbf \nu} = 1}^d 
[e^{{\mathbf g} \cdot {\mathbf e}_{\mathbf \nu}}c_{{\mathbf x} +
{\mathbf e}_{\mathbf \nu}}(t) +
e^{-{\mathbf g}\cdot{\mathbf e}_{\mathbf \nu}} c_{{\mathbf x}-
{\mathbf e}_{\mathbf \nu}}(t) - 2 cosh({\mathbf g} 
\cdot{\mathbf e}_{\mathbf \nu}) c_{\mathbf x}(t) ]
\nonumber \\
&&+  
U({\mathbf x})c_{\mathbf x}(t)-
b c_{{\mathbf x}}^2(t)\;,
\label{eq:eighteen}
\end{eqnarray}
where $c_{\mathbf x}(t)$ is the species 
population at the sites $\{{\mathbf x}\}$ of a hypercubic 
lattice with lattice constant $\ell_0$, 
and the  $\{{\mathbf e}_{\mathbf \nu}\}$ are unit lattice vectors. 
Furthermore, $w \simeq D/\ell_0^2$, where $D$ is the diffusion constant
of the corresponding continuum model,
and $g \simeq -v \ell_0/(2 D) $, where $v$ is the 
convective flow rate of the continuum model.
$U({\mathbf x})$ and $b$ have the same interpretation 
as in the continuum model  (\ref{equation-of-motion}).
The subtraction in the first term insures that 
$c_{\mathbf x}(t)$ is conserved 
($ {d\over dt}\sum_{\mathbf x}c_{\mathbf x}(t) =0 $)
if $U(x) = b = 0 $.
When linearized about $c_{\x}  \equiv 0$, 
Eq. (\ref{eq:eighteen}) may be written 
\be 
{dc_\x (t) \over dt} = \sum_{\x'}{\tilde \cL}(\x,\x') 
c_{\x'}(t)
\label{eq:nineteen}
\ee
where the discrete Liouville operator ${\tilde\cL}$ is the matrix 
\begin{eqnarray} 
\tilde \cL&=& w \sum_\x \sum_{\nu=1}^d 
[e^{-\g\cdot\e_\nu}|\x + \e_\nu\rangle \langle\x| \;  + \;
e^{\g\cdot\e_\nu}|\x\rangle \langle\x +\e_\nu | ]  \nonumber \\
&&+
\sum_\x [A +U(\x)] |\x\rangle\langle\x|
\label{eq:twenty}
\end{eqnarray}
with $A = -2w \sum_{\nu=1}^d cosh(\g\cdot\e_\nu)$.
Typical spectra with a square well growth profile
for a one dimensional 1000-site model
were shown in Fig.~\ref{Gamma-v-series} for several 
values of $g \propto v$ \cite{oasis}.
Parameter values and energies studied were chosen to give a
good approximation to the continuum.

\begin{figure}
\epsfxsize=4truein \vbox{\hskip 0truein
\epsffile{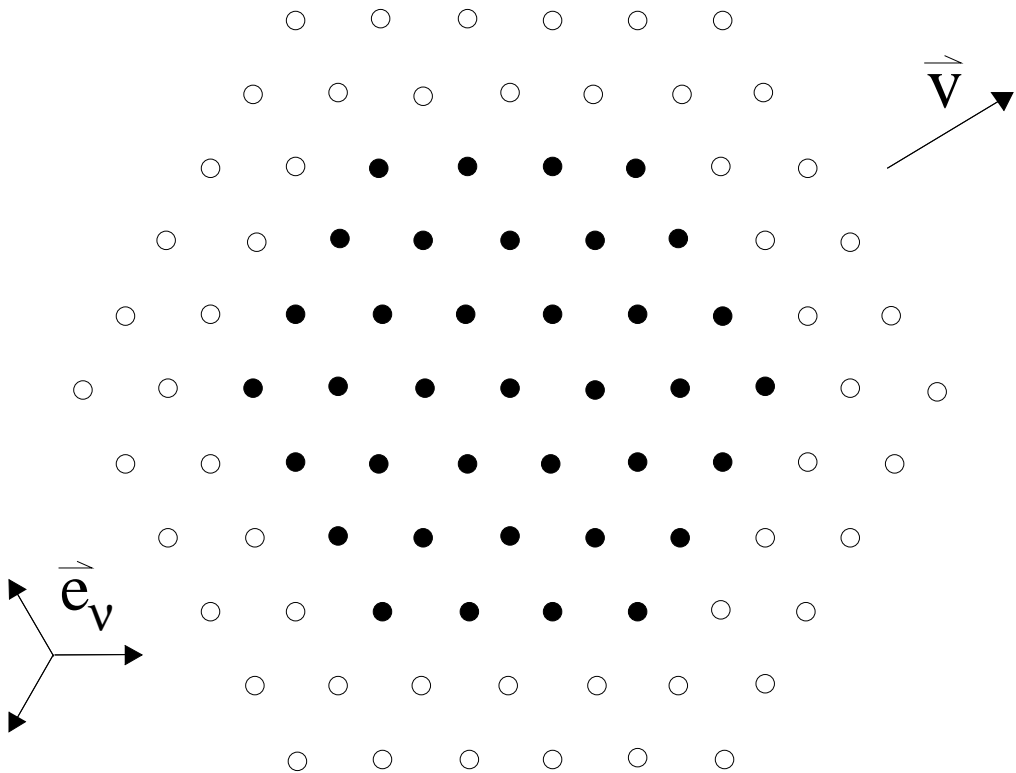}}
\vskip 0truein
\caption{Hexagonal oasis (filled circles) surrounded by 
a desert (empty circles), in 2 dimensions on a triangular lattice,
with periodic boundary conditions.
The inset in the lower left corner of the figure displays
the three basis vectors $\e_\nu$ of the lattice.
The arrow in the upper right corner of the figure indicates
the $30^{\circ}$ direction of the convection velocity 
${\mathbf v}$ relative to the horizontal axis.
(This direction renders a better approximation to the continuum
than for example a direction along one of the three lattice vectors
$\e_\nu$.)
\label{lattice-vectors}}
\end{figure}

For two dimensional systems,
a triangular lattice (see \Eq{eq:twenty-triangle}) 
is actually a better approximation to the 
continuum than a square lattice, because  
it minimizes artifacts caused
by preferred lattice directions. Using the unit lattice vectors
${\mathbf e}_1$, ${\mathbf e}_2$, ${\mathbf e}_3$ of 
Fig.~\ref{lattice-vectors} to define the Laplacian,
one finds \cite{Richtmyer} via a Taylor's series expansion
\bea
{2\over {3 \ell_0^2}} \sum_{\nu=1}^3 
[c_{{\mathbf x} + {\mathbf e}_{\mathbf \nu}}(t) + c_{{\mathbf x}-
{\mathbf e}_{\mathbf \nu}}(t) - 2 c_{\mathbf x}(t) ]
\nonumber \\
&& = \nabla^2 c_{\mathbf x}(t) + {1\over 16} \ell_0^2 
(\nabla^4 c_{\mathbf x}(t) ) + O(\ell_0^4)\, .
\eea
Using the lattice vectors ${\mathbf e}_1$, and ${\mathbf e}_2$
of a square lattice gives a similar result, but the fourth order
terms in that case are $ \ell_0^4 (\partial^4 /\partial x^4 
+ \partial^4/\partial y^4 ) c_{\mathbf x}(t)/12$ which cannot
be expressed in terms of the Laplacian of 
$c_{\mathbf x}(t)$ \cite{Richtmyer}.

\begin{figure}
\epsfxsize=4truein \vbox{\vskip 0.15truein\hskip 1truein
\epsffile{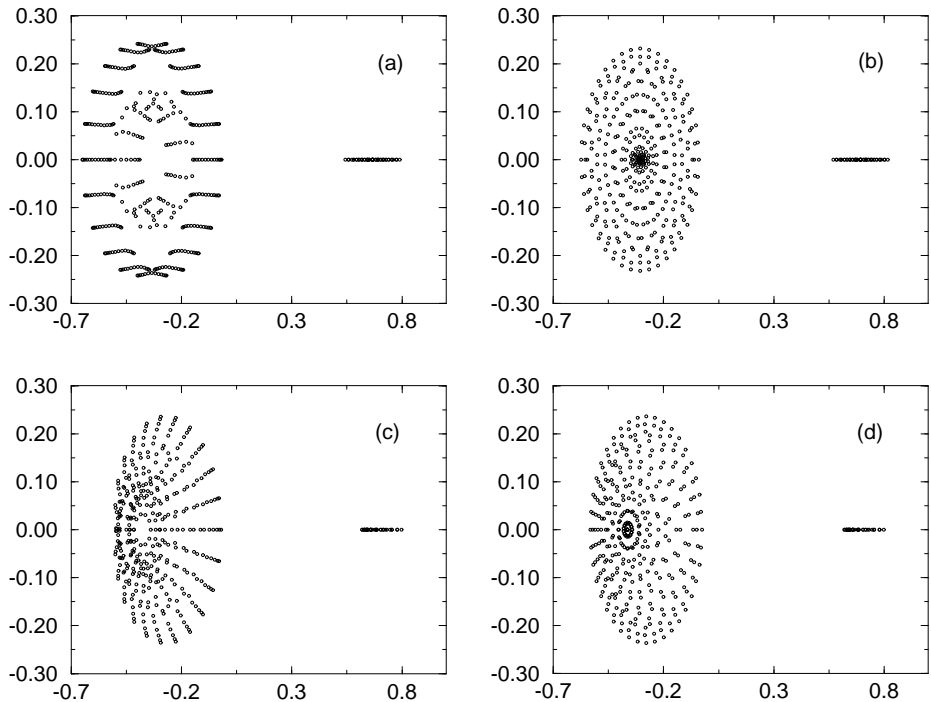}}
\medskip
\caption{Complex nonhermitian eigenvalue spectra for a two dimensional
oasis in a desert, using different lattices.
The model parameters (with $\ell_0 \equiv 1$) are
$D=w=1$, $v/v_F=0.37$, $a=29.5 w$, and 
$\epsilon a=0.5 w$.
(a) and (b) show results for a square lattice with 400 sites and
a square oasis with 36 sites, with periodic boundary conditions.
In (a) the velocity $\mathbf{v}$ is directed along the 
$x$-axis, which is one of the two preferred
lattice directions. In (b) the velocity is
directed along the diagonal ($45^{\circ}$-axis).
(c) and (d) show the results for a triangular lattice with 400 sites,
a hexagonal oasis with 37 sites, with periodic boundary conditions, as shown 
in Fig~{\protect{\ref{lattice-vectors}}}. In (c) the velocity is directed
along the $x$ axis, which coincides with the direction of one
of the lattice vectors ${\mathbf e}_{\nu}$. 
In (d) the velocity is directed along
the $30^{\circ}$-axis, as indicated in  Fig~\ref{lattice-vectors}.
(The eigenvalues are normalized by the difference of the growth rates inside
and outside the oasis $U_0=a+\epsilon a = 30 w$.)
\label{fig:lattices}}
\end{figure}

Fig. \ref{fig:lattices} shows the eigenvalue spectra at finite 
convection velocity ${\mathbf v}$ for an oasis in a desert
for two different lattices, for the square lattice, 
with ${\mathbf v}$ along the $x$ axis(a), 
and along the $45^{\circ}$-axis (b), and for the triangular lattice
(see \Eq{eq:twenty-triangle}) with
${\mathbf v}$ along the $x$ axis (c) and along the 
$30^{\circ}$-axis (d) as in Fig~\ref{lattice-vectors}.
States localized near the oasis again appear as a string of 
real eigenvalues above a pattern of complex eigenvalues representing 
delocalized eigenfunctions.
Although the ellipsoidal pattern of extended state 
eigenvalues is an artifact of the lattice, the upper edge 
can be a good approximation to the continuum if the 
direction is chosen appropriately.
Directions coinciding with ``lattice planes''
lead to spurious features
in the delocalized part of the 
spectrum. Spectral artifacts are particularly
noticeable for the square lattice, when
${\mathbf v}$ is along one of the lattice directions (a).
Note also, that for the square lattice 
the ellipse of complex eigenvalues is mirror symmetric 
with respect to the two coordinate axes through the center of the ellipse.
For neither the triangular lattice, nor the continuum model,
there is such a mirror symmetry with respect 
to the imaginary axis in the complex spectrum. 
An easy check is given by an analytic computation of 
the spectrum for a homogeneous growth rate in the system.
The results are cited in the appendix.

\begin{figure}
\epsfxsize=5truein \vbox{\vskip 0.15truein\hskip 0.5truein
\epsffile{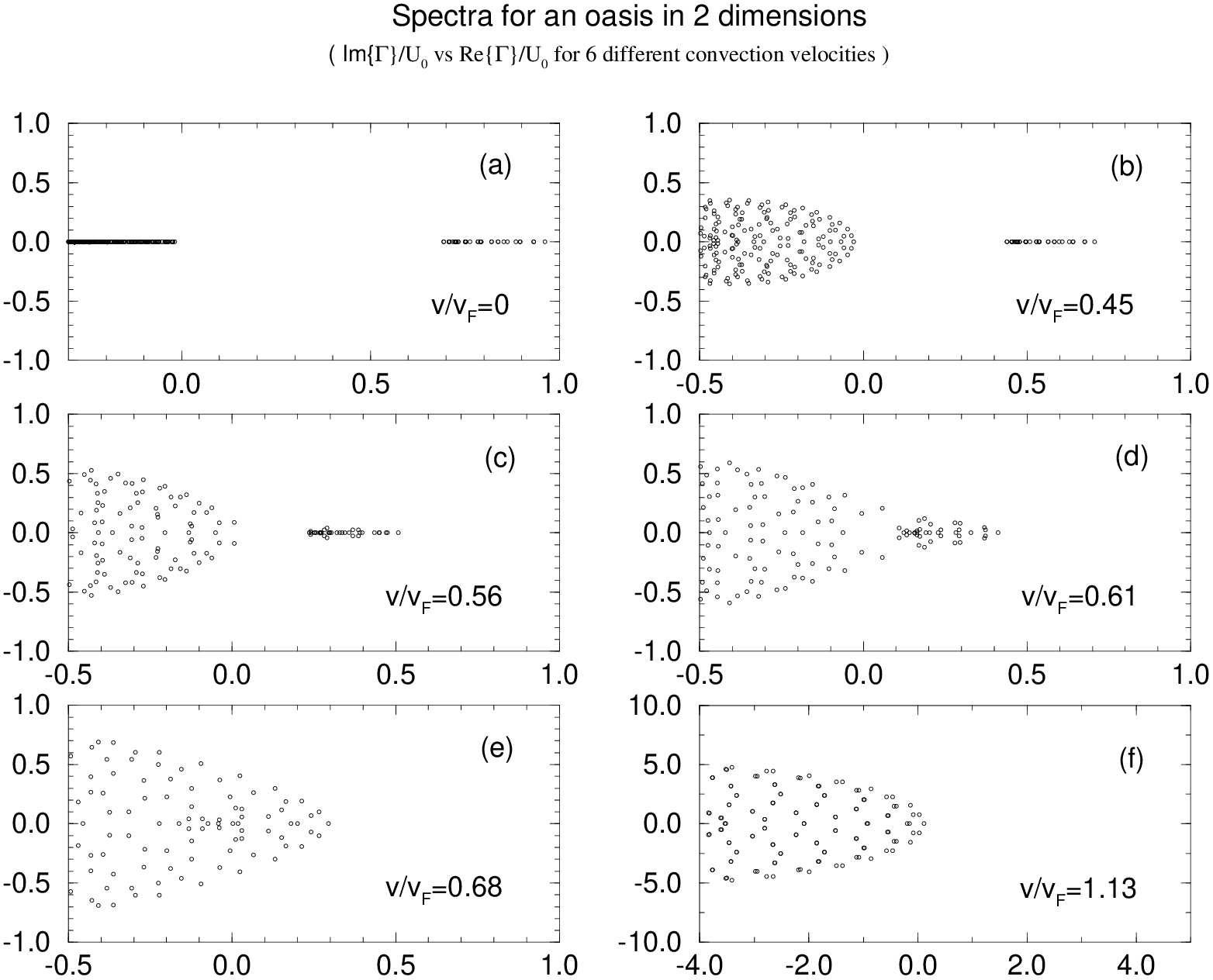}}
\medskip
\caption{Complex nonhermitian eigenvalue spectra at six different
drift velocities for a two dimensional system with a
hexagonal oasis (with growth rate $a=29.5 w'$, where $w'=1$ and
$w'\equiv 2w/3$) 
in a desert (with negative growth rate $-\epsilon a=-0.5 w'$) with periodic 
boundary conditions.
The growth rates of the spectra are divided by the difference 
of the growth rates inside and outside the oasis 
$U_0 \equiv a + \epsilon a= 30 w'$.
The drift velocities are directed along the $30^{\circ}$ axis
of the triangular 
unit cell. Their respective values 
(divided by the Fisher wave velocity inside the oasis
$v_F=2 \sqrt{a D}$) are indicated next to each figure.
The spectra are extracted from numerical simulations of the triangular
lattice model with $400$ sites, 
with a hexagonal oasis with $37$ sites, and diffusion constant 
$D=1.5 w' \ell_0^2$, with $\ell_0 = 1$.
(This figure actually only shows the part of the spectrum 
which gives a good approximation to the continuum problem.) 
\label{2-dim-spectra}}
\end{figure}

We now use a triangular lattice to study the delocalization transition
in two dimensions. For a triangular lattice the sum
in \Eq{eq:twenty} over $\nu$ is replaced by a sum over three 
unit lattice vectors illustrated in Fig~\ref{lattice-vectors}.
One obtains
\begin{eqnarray} 
\tilde \cL&=&  {2w \over 3} \sum_\x \sum_{\nu=1}^3 
[e^{-\g\cdot\e_\nu}|\x + \e_\nu\rangle \langle\x| \;  + \;
e^{\g\cdot\e_\nu}|\x\rangle \langle\x +\e_\nu | ]  \nonumber \\
&&+
\sum_\x [A+U(\x)] |\x\rangle\langle\x|
\label{eq:twenty-triangle}
\end{eqnarray}
with $A = -{4w\over 3} \sum_{\nu=1}^3 cosh(\g\cdot\e_\nu)$.
Fig.~\ref{2-dim-spectra} shows lattice simulation results for
six spectra at different convection
velocities for the linearized two dimensional system 
with a hexagonal oasis, and periodic
boundary conditions. The convection velocity is
oriented along the $30^{\circ}$ axis relative to the lattice,
as shown in figure 
\ref{lattice-vectors}. 
At zero velocity all eigenvalues are real 
(figure \ref{2-dim-spectra}(a)). Just as 
in one dimension, with increasing 
convection velocity, the real growth rates of the localized states 
decrease while the mobility edge ({\it i.e.} the rightmost eigenvalue
of the parabolic envelope of the
delocalized states with complex eigenvalues) remains unchanged (b).
Unlike in the one dimensional case, 
there are delocalized states with similar real parts of their eigenvalues,
but a spread in the imaginary parts.
(The reason is that in two and higher dimensions there are delocalized
states that can carry a quantum mechanical current in directions other 
than the convection velocity \cite{Hatano,Brouwer}.)
Also, as the convection velocity
is increased, some of the localized states become delocalized
(signaled by eigenvalues which escape into the complex plane \cite{Hatano})
even before joining the parabola of delocalized states (c).
The parabola formed by the eigenvalues of the delocalized 
states widens as
the imaginary parts of the eigenvalues grow with the convection velocity,
just as we saw for one dimensional systems.
For even higher convection velocity the eigenvalues of the localized
states enter the parabola of the delocalized states. The mobility 
edge is ill defined at these intermediate velocities (d).
As the convection velocity $v$ approaches the Fisher wave velocity 
$v_F = 2 \sqrt{a D}$,
the upper edge of the band settles again into its fixed value 
(approximately
$Re\{\Gamma\} =0$ for the given parameter values), 
where it remains for all higher velocities (f).

\begin{figure}
\epsfxsize=4truein \vbox{\vskip 0.4truein \hskip 1.1truein 
\epsffile{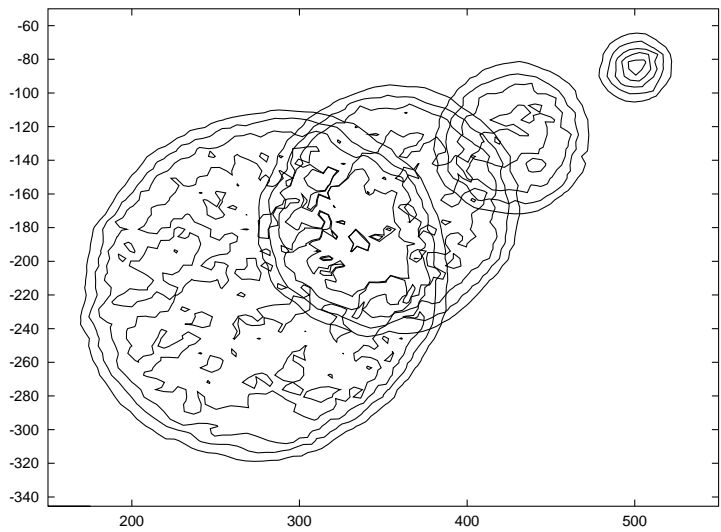}}
\medskip
\caption{Contour plots for snapshots of the distribution of bacteria in a 
two dimensional landscape with (quenched) random growth rates,
obtained from the full nonlinear 
problem for convection in the direction relative to a 
triangular lattice shown in Fig.~\ref{lattice-vectors},
with velocity parameter $g=1$, $w' \equiv 2w/3 =1 $, $b=1$, and $\ell_0=1$. 
The random growth rates $U(\mathbf{x})$ are uniformly distributed
with values ranging from $-0.3 w'$ to $0.7 w'$.
At time t=0 the bacterial population was zero everywhere, except
at one lattice point where it was set to 1.
The snapshots show the expanding and drifting droplet
after $250$, $500$, $750$, and $1000$ time steps of size $\delta t=0.1/w'$.
The contour lines marking population density $c=0.1$ are the innermost
contour for the snapshot after 250 time steps, the second to outermost 
contour for the snapshots after 500 and 750 time steps and the outermost
contour for the snapshot after 1000 time steps. 
The simulated system had $500 \times 500$ sites.
\label{snapshot}}
\end{figure}

{\it Dynamical} properties of the full nonlinear system can be studied
by discretizing \Eq{eq:nineteen} also in time and 
using Runge Kutta methods to solve the equation numerically
\cite{mathematica,Ames,Richtmyer,numericalrecipes}. 
Although we hope to present a more thorough study at a later date,
Fig.~\ref{snapshot} shows preliminary results in the form of
snapshots of a bacterial colony drifting and expanding 
in a 2 dimensional system on a triangular mesh,
with convection and quenched random growth rates. 
The initial condition for the system
shown was a delta function like spatial distribution concentrated
in the upper right hand corner.
Theoretical predictions in the limit of high convection velocity
for the long time scaling behavior
of a growing and
drifting droplet of bacteria in the linearized regime
were given in \cite{nadav}.

\section{Exact Results for Lattice Models in One Dimension}
\label{sec:IV}

Consider the lattice models discussed in the previous section with a
{\it random} growth rate $U(\mathbf x)$  chosen from an identical proability
distribution at every lattice site $\mathbf{x}$. If the drift parameter
$\mathbf{g} \propto \mathbf{v}$ 
in \Eq{eq:twenty} is set to zero, the growth operator 
$\tilde \cL$ becomes formally identical to a Hermitian Hamiltonian
describing a quantum mechanical particle which hops between lattice 
sites in a disordered medium. A number of exact results are known
for this problem in one dimension; we review here recent exact results
for the non-Hermitian generalization embodied in \Eq{eq:twenty}
\cite{brouwer2,brezin,russim}, with an emphasis on the winding numbers 
and complex currents associated with extended states
\cite{Shnerb-Nelson-PRL}.

A numerical study of the eigenvalue spectrum of the lattice model in 
one dimension \cite{Hatano} reveals the following: when $g=0$,
{\it all} eigenfunctions $\psi_n(x)$ are localized with a bounded
point-like spectrum $\{\Gamma_n\}$ of eigenvalues lying on the real axis,
similar to the {\it subset} of localized states which fall to the right 
of the mobility edge for the square well spectrum shown in 
Fig.\ref{Gamma-v-series}. Localization of all states by a random 
potential is the expected result for quantum particles in a disordered 
medium in one dimension. The situation changes dramatically, however,
when $g \neq 0$. For $g$ greater than a critical value $g=g_c$,
a bubble of complex eigenvalues appears in the center of the band, similar 
to the response to convection for arbitrary $g$
of the continuous spectrum part of the 
spectrum for a square well shown in Fig.\ref{Gamma-v-series}.
The eigenfunctions associated with this bubble are {\it extended}
\cite{Hatano}. Eigenvalues outside this bubble remain real, and are
related by a simple rigid band shift to their values for $g=0$.
The right and left eigenfunctions $\phi_n^{R,L}(x)$ for states
outside the bubble remain localized, and are related to the eigenfunctions
$\psi_n(x_j)$ of the Hermitian problem for $g=0$  by 
\be
\label{gauge-right-left}
\phi_n^{R,L}(x_j) = e^{\pm g x_j} \psi_n(x_j)\, .
\ee
We assume here a one dimensional lattice with unit spacing $\ell_0=1$.

We shall focus on the
eigenfunctions and complex currents associated with 
the band of extended states in one dimension for lattice growth 
models with disorder in one dimension \cite{Shnerb-Nelson-PRL}.
Unlike delocalized states in Hermitian disordered systems (where the
eigenfunctions can always be chosen to be real), we show that these 
complex eigenfunctions 
are characterized by a conserved  winding number $n$, 
even when the disorder is 
 strong. Such topological quantum numbers 
can be used to label the eigenvalue spectrum $\Gamma_n(g)$,
where $g$ is the
asymmetry parameter.
A study of the eigenvalue trajectories as a function of 
$g$ then leads to complex currents defined by
$J_n = -i {\partial \Gamma_n \over \partial g}$, which determine
the response of the $n$-th eigenvalue to changes in the drift velocity. 

The  matrix representation of the   one dimensional non-Hermitian  
Liouville operator (\ref{eq:twenty}) in a basis of 
$N$  sites localized at positions  $\{ x_j, j=1, \cdots, N  \} $  reads,
\be 
\label{matrix-1}
\tilde \L  =  \left( 
\begin{array}{ccccc}
U_1 & {w_1 \over 2} e^{g_1}  &0    &..&{w_N \over 2}   e^{-  g_N }  \\
{w_1 \over 2} e^{-g_1}  & U_2 &{w_2 \over 2}e^{g_2}   &..&0 \\
0& {w_2 \over 2} e^{-g_2}  &U_3&..&0 \\
:&:&:&:&: \\
{w_N \over 2}  e^{ g_N }    &0&0&.. &U_N \\
\end{array}
\right). 
\ee   
where we assume periodic boundary conditions, and
the constant background of the  growth  rate $A$ has been omitted. 

The matrix displayed in \Eq{matrix-1} (tridiagonal, with nonzero corner
matrix elements) is actually a generalization 
of \Eq{eq:twenty}: As before, $U_j$
is a zero mean random growth rate chosen independently for each
site. However, the hopping rate between sites $i$ and $i+1$,
$w_i$, is now a random function of position, which allows for 
fluctuations in the local diffusion constant. We have also allowed for 
fluctuations in the local convection velocity $g_i$. This latter variation
can be eliminated via similarity
transformation which preserves the eigenvalues of $\tilde \L$,
and produces only a smooth, nonsingular change in the eigenfunctions.
It is helpful to carry out this transformation in two stages. The 
first isolates all effects of convection on the bond connecting
the $N$th and 1st site: Upon defining
\be 
\label{matrix-S1}
S_1  =  \left( 
\begin{array}{ccccc}
1 & \quad 0  & \;\; 0    &..&0  \\
0  & \;\quad e^{-g_1} & \;\; 0   &..&0 \\
0& \quad 0  &\;\;\;e^{-g_1-g_2}&..&0 \\
:&\quad :&\;\; :&:&: \\
0   & \quad 0&\;\;0&.. &\; e^{-\sum_{j=1}^{N-1} g_j} \\
\end{array}
\right) 
\ee   
we have
\be 
\label{matrix-Lprime}
\L'= S_1^{-1} \tilde{\L} S_1 =  \left( 
\begin{array}{ccccc}
U_1 & \;\;{w_1 \over 2}  & \quad 0    &\quad ..&{w_N \over 2}   e^{- \sum_{j=1}^N g_j }  \\
{w_1 \over 2} &\;\; U_2 & \quad {w_2 \over 2}   &\quad ..&0 \\
0& \;\;{w_2 \over 2}  &\quad U_3&\quad ..&0 \\
:&\;\; :&\quad :&\quad :&: \\
{w_N \over 2}  e^{\sum_{j=1}^N g_j }    &\;\; 0&\quad 0&\quad .. &U_N \\
\end{array}
\right). 
\ee   
Except for the corner matrix elements, $\L'$ describes {\it Hermitian}
population growth without convection. A second similarity transformation,
\be 
\label{matrix-S2}
S_2  =  \left( 
\begin{array}{ccccc}
1 & 0  &0    &..&0  \\
0  & e^{g} & 0   &..&0 \\
0& 0  &e^{2 g}&..&0 \\
:&:&:&:&: \\
0   &0&0&.. &e^{(N-1) g} \\
\end{array}
\right) 
\ee  
where $g={1\over N} \sum_{j=1}^N$ then redistributes the convection
{\it uniformly} on all bonds,
\be 
\label{matrix-L}
\L= S_2^{-1} \L' S_2 = S_2^{-1} S_1^{-1} \tilde{\L} S_1 S_2 =  
\left( 
\begin{array}{ccccc}
U_1 & {w_1 \over 2} e^g & 0    &..& {w_N \over 2} e^{-g}   \\
{w_1 \over 2} e^{-g}  & U_2 & {w_2 \over 2} e^g   &..&0 \\
0& {w_2 \over 2} e^{-g}  &U_3&..&0 \\
:&:&:&:&: \\
{w_N \over 2} e^g    &0&0&.. &U_N \\
\end{array}
\right). 
\ee   
The composite similarity matrix
\be 
\label{composite-matrix}
S_{ij} = (S_1 S_2)_{ij} = \delta_{ij} \exp[-\sum_{k=1}^{j-1} g_k + 
{(j-1)\over N} \sum_{k=1}^N g_k]
\ee
thus leads to a population growth problem whose eigenvalues
reflect a mean convection parameter $g={1\over N}\sum_{j=1}^N g_j$.
For the $g_j$'s chosen independently at each site,
the sample to sample
fluctuations should fall off like $\sigma/N$ where $\sigma$
is the variance of the distribution.
As $N$ grows one may thus  replace 
the fluctuating quantity $g_i$ by a disorder independent average value 
 $g$, as we shall 
 do in the rest of this 
paper. This result implies 
{\it universality} in the response to random convection
in one dimensional growth models - nonuniform convection velocities may be mapped 
into a uniform average velocity via a similarity
transformation,
provided one applies the transformation~(\ref{composite-matrix})
to the eigenfunctions as well. 

For a 1D ring with random parameters $\{U_i\} , \{w_i\}$ and    $g = 0$,
all the eigenfunctions of (\ref{matrix-Lprime})  are real and localized, 
and its eigenvalues are real and discrete \cite{Hatano,nadav}. 
We assume for simplicity that all $w_i >0$ and that the chain is 
large but 
finite, such that although the spectrum is discrete,  the 
length of the chain is much larger than the maximal localization length
of an eigenmode.

\begin{figure}
\epsfxsize=3.5in \vbox{\vskip 0.2truein \hskip 1.3truein 
\epsffile{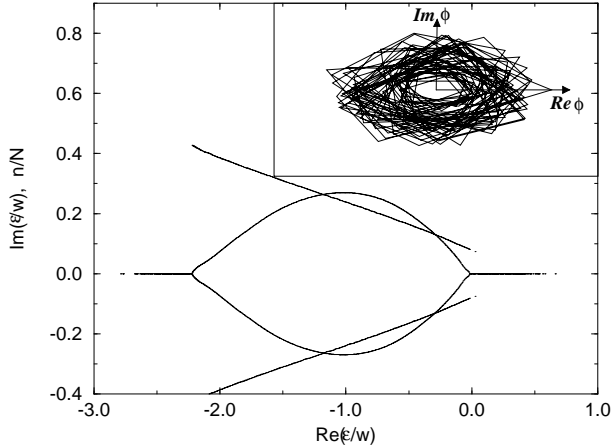}}
\medskip
\caption{Spectrum and  winding numbers 
for  $N = 1000$  with $U_j$ uniformly distributed 
in the interval $[-1,1]$,  and $g=0.4$. 
Inset: Projection onto the complex plane
of an extended eigenfunction near the center of the bubble
with $N = 200$, $g=1$, and winding number $n=100$.
\label{winding-numbers}}
\end{figure}

In a typical one-dimensional  disordered system
with ${g} = 0 $, the localization 
length $\xi$ (defined via the exponential fall off of the eigenfunction)
is larger at the center of the band and smaller at the tails.
The criterion  for delocalization of the asymmetric system
via convection (see \Eq{gauge-right-left}) is $\kappa \ell_0 < g$, 
where $\kappa \equiv 1/\xi$ \cite{Hatano}.  
As a result, pairs of complex energies representing  delocalized states
first appear as a ``bubble'' at the center of the band, 
which  then spreads into the band tails as the convection parameter
$g$ is increased (see Fig.~\ref{winding-numbers}).  
To study the complex currents associated with delocalized states 
we  follow Refs.\cite{brouwer2,brezin,russim} and exploit the  
relation between the complex spectrum of the  asymmetric problem
 with $g \neq 0$  and the real eigenvalues of a 
``background'' localized problem with  $g=0$, with the   
same realization of disorder.
\begin{figure}
\epsfxsize=3.2in \vbox{\vskip 0.2truein \hskip 1.3truein 
\epsffile{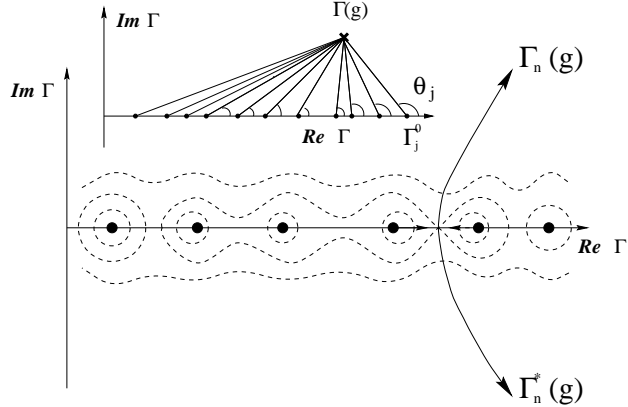}}
\medskip
\caption{Eigenvalue trajectories in the 
complex plane. Points on the real axis are eigenvalues for the 
Hermitian problem for $g=0$. The complex eigenvalue trajectories 
$\Gamma_n(g)$ and $\Gamma_n^*(g)$ are orthogonal to the
level curves (dashed lines) defined by Eq.(\protect{\ref{loc}}).
Inset: Angles entering Eq.(\protect{\ref{arg}})
for a given eigenvalue $\Gamma(g)$.  
\label{evtrajectories}}
\end{figure}
The condition for a complex number $ \Gamma  = \Gamma_R + i\Gamma_I $ 
to be an eigenvalue of the matrix $\L(g)$ displayed in \Eq{matrix-L}
is $Det[\Gamma I -\L(g)] = 0$,
where $I$ is the identity matrix. 
Upon expanding in cofactors, we find that $\Gamma$ is an eigenvalue 
provided \cite{brouwer2,thouless} 
\begin{eqnarray} \label{thou}
Det[\Gamma I -\L(g=0)] &=& \prod_{i=1}^N (\Gamma - \Gamma^0_i) \nonumber \\
 &=&  
   2 [\cosh(g N) - 1]  \prod_{j=1}^N  \left( {w_j \over 2} \right)   
\end{eqnarray}
where the  $\{\Gamma^0_i\}$ 
are the (real) eigenvalues of the background matrix  $\L(g=0)$. 
To extract winding numbers, we first observe 
that  the right hand side 
of \Eq{thou} is real and positive. As a result,   
the phases of the complex numbers in the left product of (\ref{thou})
for
each complex $\Gamma$ 
should sum up to
(see inset to Fig.~\ref{evtrajectories})
\begin{eqnarray} \label{arg}
\sum_i cot^{-1} ({\Gamma_R - \Gamma^0_i \over \Gamma_I}) = p \pi,
\end{eqnarray}
with  $p = 2n$, where $n$ is an integer, and the function 
$cot^{-1}(x)$  varies from $\pi$ to $0$ as 
$x$ goes between $-\infty$ and $\infty$.

As $\Gamma_I \to 0$,  each term in Eq.(\ref{arg})   
gives $\pi$ for every  eigenvalue $\Gamma_i^0$ 
 to the right of $\Gamma_R$, and zero 
for each eigenvalue to the left (see Fig.\ref{evtrajectories}). 
To satisfy (\ref{arg}), 
the eigenvalue must leave the real axis and enter  the complex plane 
at the gap between the 2n-th and the 2n+1-th 
eigenvalues  of the 
$g=0$ ``background'' system.   We call $n$ the {\it index} of the 
trajectory $\Gamma_R(g) + i \Gamma_I(g) $ in the complex plane.

We see immediately from (\ref{arg}) that the rightmost 
eigenvalue (with  $n=0$) must remain real, consistent 
with Perron-Frobenius theorem  \cite{peron}. 
The corresponding nodeless eigenfunction corresponds to 
the ground state of $ -\L'$. For $N$ even, particle-hole symmetry 
implies that {\it both} the rightmost and leftmost eigenvalues are 
always real. 
More generally, for a fixed value of $n$, the set of all 
$[\Gamma_R(g), \Gamma_I(g)]$ satisfying Eq. (\ref{arg}) 
defines a curve in the complex plane, as 
illustrated in Fig.\ref{evtrajectories}. 
Henceforth, we assume $N$ even for simplicity.  

A more  complete description of the eigenvalue trajectory
results from 
taking the logarithm of the modulus  of Eq. (\ref{thou}).
In  the limit $N g >>1$ one finds
a second  constraint on $\Gamma(g) $, namely, 
\cite{brouwer2,brezin,russim}
\be \label{loc}
|g| = -\ln({w \over 2}) + {1 \over N} \sum_i \ln (|\Gamma - \Gamma^0_i|) 
\ee
where $|\Gamma - \Gamma^0_i|  = \sqrt{(\Gamma_R-\Gamma^0_i)^2 
+ \Gamma_I^2}$ and $w = 2   
[\prod_{j=1}^N {w_j \over 2} ]^{1 \over N}$. 

This constraint is described graphically in 
Fig. \ref{evtrajectories}, which shows schematically 
the level curves defined by \Eq{loc} 
near the band center for three values of $g$. 
These are lines 
of constant potential for an equivalent 2d electrostatic problem with 
charges at the positions of the localized point spectrum for $g=0$. 
When $g$ is small, the constraint is solved by eigenvalue pairs on the 
real axis in the gaps between 
neighboring $\Gamma^0_j$'s indexed by $p=2n$. 
As $g$ increases,
successive pairs of eigenvalues eventually merge at a saddle point 
in the potential contours  
and detach from the real axis at right angle. For a given real  energy 
$\Gamma_R$,  Thouless has defined an energy dependent inverse localization 
length $\kappa(\Gamma_R)$ for the associated $g=0$
Hermitian problem \cite{thouless},
 \be
\kappa(\Gamma_R)  
= -\ln({w \over 2}) + {1 \over N} \sum_i \ln (|\Gamma_R - \Gamma^0_i|).
\ee
Upon comparing with Eq.~(\ref{loc}), we see that $\Gamma_I$ becomes 
nonzero 
whenever $|g| > \kappa(\Gamma_R')$ where $(\Gamma_R', 0)$ is the 
detachment point of the eigenvalue pair. 

As $g$ increases above $\kappa(\Gamma')$, Eqs.~(\ref{arg}) and (\ref{loc}) 
define a unique pair of 
complex eigenvalue trajectories $\Gamma_n(g)$ and $\Gamma_n^*(g)$
for every value of $n$. 
Upon passing to the limit $N \to \infty$, the spectrum $\{ \Gamma_j^0 \}$
for $g=0$ closes up, and is described by a density of states 
$\rho_0(\lambda)$.
Eqs.~(\ref{arg}) and (\ref{loc}) 
may then be recombined into a single complex equation, namely

\be\label{8}
\int_{-\infty}^{\infty} d \lambda \; \rho_0(\lambda) \ln[\Gamma_n (g)  
- \lambda]
 = \ln \left( {w \over 2} e^{|g|} \right)  + 
i \pi \left( {2 n \over N} \right).
\ee
where 
\be
\ln[\Gamma (g)  - \lambda] = \ln[|\Gamma_R +i \Gamma_I   - \lambda|] + i cot^{-1}
[(\Gamma_R-\lambda)/\Gamma_I].
\ee
For simplicity we set the average growth rate to zero. Then in
the limit $N >> 1$ and for  a density of states function 
symmetric around the special detachment point with $\Gamma_R' = 0$, 
there is a purely imaginary trajectory of the form $\Gamma(g) 
= i \Gamma_I(g)$ with $n = N/2$. For fixed $g$, Eq.~(\ref{8}) thus  
leads to an implicit formula 
for the ``height'' $\Gamma_I^{max}$ of the bubble of 
complex eigenvalues in the 
center of the band (see Fig.~\ref{winding-numbers}), namely,
\be
{1 \over 2} \int d \lambda \;  \rho(\lambda) \;   
\ln [\lambda^2 + (\Gamma_I^{max})^2] =
|g|   + \ln({w \over 2}).
\ee
This integral vanishes, as expected \cite{brezin},  
in the ``one way'' limit, $g \to \infty$ with $ {w \over 2} e^{|g|} = 1$.
For other detachment points, the eigenvalue trajectories 
curve to the left or right as required by the constraint of \Eq{loc} 
(see Fig.~\ref{evtrajectories}). 

The analysis above  
suggests that the imaginary parts of all $\{ \Gamma_n(g) \}$
(except the two which remain on the real axis),
diverge as $|g| \to \infty$. 
It is then expected that  {\it all}
eigenfunctions $\phi_n(j)$ are approximately 
plane waves, $\phi_n(j) \sim \exp(ik_n r_j /\ell_0)$, with free particle 
eigenvalue 
energies $\Gamma_n(g) = w \cos(k_n + ig)$ \cite{nadav,Hatano}. 
For large 
$|\Gamma_I(g)|$,  Eq. (\ref{8}) leads to
\be
\Gamma_n(g) \approx  {w \over 2} \exp(|g| + 2 i \pi n/ N).
\ee
Comparison with the free particle spectrum at large $|g|$ shows immediately
  that the index $n$ of the eigenvalue trajectory and the wave vector are 
related, $k_n = 2  \pi n / N$. These  wave eigenfunctions spiral around the 
origin in the complex plane as one moves along the 1d lattice of the tight 
binding model  sites, leading to a well defined winding number $n$. 
The winding number associated with the eigenvalue 
$ \Gamma^*_n(g)$ in the lower half plane is then $-n$. 

As $g$ decreases, it can be shown that 
the associated delocalized wave function $\phi_n(j;g) $
must {\it remain} nonzero for all $j$, so long as 
the eigenvalue retains an imaginary part \cite{comment}.
Thus, the  winding number must be {\it preserved} as $g$ 
decreases, i.e., the winding number is a topological invariant along an 
eigenvalue trajectory. The projection of such a 
delocalized eigenfunction is illustrated in the inset to  
Fig.~\ref{winding-numbers}.

As $N \to \infty$, we can replace the winding index $n$ by the continuous
variable $k_n = {2 \pi \over N} n$. Eq. (\ref{8}) then shows quite generally 
that the complex spectrum is a function only of the combination 
$g+ik$, i.e., $\Gamma_n(g) = \Gamma(g+ik)$. It then follows from the 
Cauchy-Riemann relations that all eigenvalue trajectories are at 
right angles to the lines of constant $g$, {\it i. e.}
${d \Gamma(g) \over dg} = -i 
{d \Gamma(g) \over dk}$. 
An explicit formula for the complex current 
results, moreover, from differentiating Eq. (\ref{8}) with respect to $g$,
\begin{eqnarray}\label{11}
J[\Gamma_n(g)] &=& -i {\partial  \Gamma_n(g) \over \partial g}
\nonumber \\ &=& 
\left[ i  \int d\lambda \; { \rho_0(\lambda)  \over \Gamma_n(g) - \lambda}  
\right]^{-1} \equiv - i G_0^{-1} [\Gamma_n(g)]
\end{eqnarray}

Evidently, the current associated with a particular complex eigenvalue 
$\Gamma_n(g)$ is determined by the Green's  function of the 
$g=0$ problem, evaluated at the complex energy $\Gamma_n(g)$. 

Fig.~\ref{winding-numbers} 
shows a typical eigenvalue spectrum for $\L$ in one dimension  
superimposed  on the winding numbers associated with the extended 
eigenfunctions. The inset illustrates how the winding number is defined
for a typical extended state. The winding numbers are exactly 
$ \pm N/2$ at the band center, and their magnitudes decrease
monotonically as one moves toward the upper edge of the band, i.e., toward 
the lowest energies of the corresponding Hamiltonian. The winding numbers 
remain finite  up to the mobility edge separating complex and real 
eigenvalues, but become undefined in the band tails, where all wave functions 
are real. At the transition, an overall constant phase factor can be adjusted
so that the imaginary part of the eigenvector
vanishes while its real part remains finite. In this case, 
the winding index classification 
can be replaced by counting the nodes of  the real, localized wave function 
\cite{thouless}.

\section{Pseudospectra and Sensitivity to Boundary Conditions}
\label{sec:V}

In this review, we have explored the spectral properties of the 
linearized  growth operator $\L$ 
in order to learn about the time evolution of the growing populations
using spectral decomposition, as in \Eq{time-evolution}. 
In the recent years, however, it has been recognized
that such a straightforward analysis 
of non-Hermitian operators and matrices (such as the Liouville operator
in \Eq{Liouville}) can in some circumstances lead to inaccurate
predictions \cite{trefethen1,trefethen2,Reddy}.
In this section we discuss convection--diffusion
with a random growth rate in this context. 
We argue that, while some non-Hermitian problems are indeed
abnormally sensitive to small perturbations, our results
(provided the boundary conditions admit a net population flux)
are quite robust. We consider a linear, 
(possibly non-Hermitian),
operator which admits matrix representation in a Hilbert space.
When convenient, we interchange freely 
between the matrix and the operator 
descriptions.

Consider first the {\it  Hermitian} operator, $\H$, 
 which   has a  complete set
of orthogonal eigenvectors. If the spectrum of $\H$ is non-degenerate, 
any eigenvalue $\Gamma_n$
corresponds uniquely to an eigenstate. When a two-fold degeneracy 
occurs, there are two independent eigenvectors corresponding to the 
same eigenvalue, and one may orthogonalize these states using the 
Graham-Schmidt scheme \cite{peron}. 
If $\H$ is an evolution operator,
its  response to external perturbation with ``frequency'' $z$ is determined 
by the resolvent ({\it i.e.} the Green's function) 
$(z I - \H)^{-1}$. The resolvent norm
\cite{normdef} diverges if $z$ is a spectral point of $\H$. For small
deviations, i.e., $z = \Gamma_n  +  \delta z$, for some $\Gamma_n$
one may use the diagonalized 
representation of the resolvent to see that its norm is proportional 
to $(\delta z)^{-1}$. 

For a non-Hermitian operator, or, in particular, a real asymmetric 
object, such as $\L(g)$, 
the situation is different. If the spectrum of this non-normal   
operator is non-degenerate, it 
admits {\it two} complete, {\it biorthogonal} states of left and right 
eigenfunctions \cite{Hatano98}. 
For any eigenvalue $\Gamma_n(g)$ the left eigenstate $<\phi^n_L|$
satisfies $<\phi^n_L| \L(g) = \Gamma_n(g) <\phi^n_L|  $ while the right state
satisfies $ \L(g) |\phi^n_R > =  \Gamma_n(g) |\phi^n_R >$. The biorthogonality 
condition means that
\be \label{norm}
<\phi^m_L|\phi^n_R >  = \delta_{m,n}.
\ee
Unlike the Hermitian case, the left and right 
eigenstates are {\it not} simple complex conjugates.
Hence, the normalization condition (\ref{norm}) implies no restriction on
the usual Hilbert norm of the left and right eigenfunction considered 
separately, and the only normalized 
quantity is 
the ``overlap integral'' between them. In the following we 
explore the significance of this observation. 

\begin{figure}
\epsfxsize=3.2in \vbox{\vskip 0.2truein \hskip 1.3truein 
\epsffile{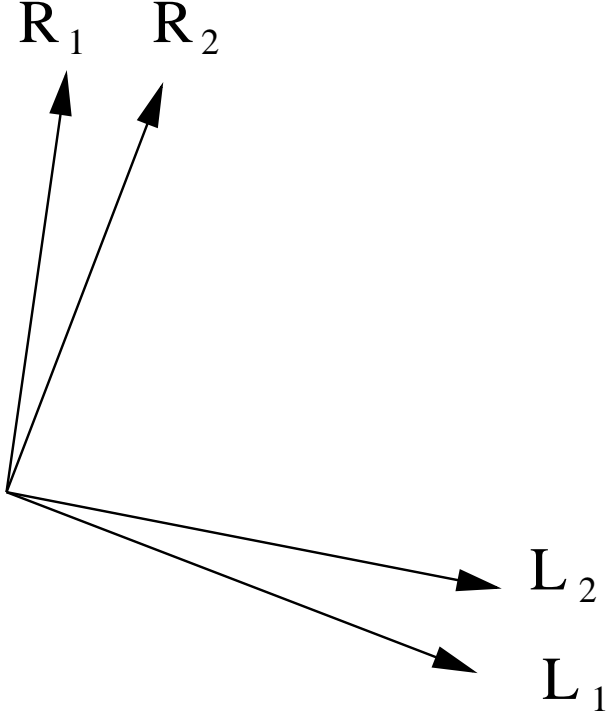}}
\medskip
\caption{Two pairs of almost degenerate,  left and right  eigenvectors of a 
non-normal 
matrix. The  condition (\protect{\ref{norm}}) implies 
that the $R_1$ and $L_2$ are orthogonal, as well as $R_2$ and $L_1$. Close to 
degeneracy $L_1$ and $L_2$ almost coincide, hence $R_1$ is 
``almost orthogonal'' to $L_1$. As illustrated in Fig.~\ref{overlap},
a typical value of the poperly normalized eigenfunctions must then be very 
large in order to satisfy
the normalization condition  $<L_n|R_n>  = 1$.
\label{deg}}
\end{figure}

Non-normality also implies that, 
in the generic case, as two eigenstates coincide (when the spectrum is 
``degenerate'') the corresponding  eigenvectors also coincide. The
subspace spanned by these vectors is one dimensional, and an $N \times N$ 
matrix is {\it defective}, i.e., it  fails
to span the $N$ dimensional  vector space associated with it. An illustration 
of the ``almost degenerate'' situation for the left and right  
eigenvectors in a $2 \times 2$ space is represented in Fig.~\ref{deg}.
Note that $ < L_2 | $ is orthogonal to $ | R_1 > $, and $ < L_1 | $
is orthogonal to $ | R_2 > $ as required by \Eq{norm}.

  The notion of pseudospectra has been introduced
\cite{trefethen1} in order
to quantify the problems which may arise when spectral
analysis is applied to non-normal operators. 
While the spectrum $\Gamma(\L)$
of the  operator $\L$ is defined
as the set of points (in the complex plane) which are eigenvalues 
of $\L$, the {\it $\epsilon$-pseudospectrum}  $\Gamma_\epsilon (\L)$
is defined as a set of points which imitate a true 
eigenvalue to within a small parameter $ \epsilon > 0$. Formally,  
the definition is:

\be \label{def1}
\Gamma_\epsilon (\L) = \{ complex \  numbers  \ z \  such \  that \ 
 ||(zI - \L)^{-1}||
 \geq  \epsilon^{-1} \}
\ee
(note that if $z$ is a spectral point, $ ( ||zI - \L||)^{-1} = \infty$). 
Eq.~(\ref{def1}) implies that the response of the system to a small
time dependent 
perturbation with ``frequency'' $z$ mimics  
a resonance up to times of order $ t \sim 1/\epsilon$. If the 
pseudospectral region is large for small $\epsilon$, the system is highly 
unstable against weak external perturbations even
far away from the resonance. 

Another definition (mathematically equivalent to (\ref{def1}))
is,
\bea \label{def2}
\nonumber
\Gamma_\epsilon (\L) &=& \{ complex \  numbers \  z \  which \  belong \
to \  
the  \ spectrum \  of   \ 
\L + \delta \L,  \\
& & for \ all \ \delta \L \ such \ that \ ||\delta \L|| \leq \epsilon \}.
\eea
Eq.~(\ref{def2}) defines the pseudospectrum via the sensitivity of 
$\L$  to small perturbations in the growth operator. Again,
broad pseudospectral regions for small $\epsilon$ imply high sensitivity
to small perturbations.

\begin{figure}
\epsfxsize=3.2in \vbox{\vskip 0.2truein \hskip 1.3truein 
\epsffile{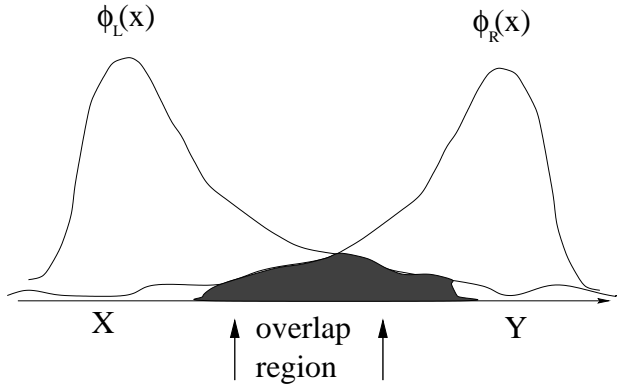}}
\medskip
\caption{Sketch of the right and left eigenfunctions of a  non-normal operator.
Since the normalization is given by the overlap  (shaded) region,
the spectrum of such a system is very unstable against  
a non-diagonal perturbation which connects the regions $X$ and $Y$.
\label{overlap}}
\end{figure}

The reason for such an extreme sensitivity in the context of the random
growth operator considered here, can be clarified if one looks 
at the (localized) left and right states shown in Fig.~\ref{overlap}. 
For simplicity, we show the ``ground state'' (most rapidly growing)
left and right eigenfunctions, but similar considerations apply to
excited states as well.
For nonzero convection velocities the peaks in the eigenfunction will be 
shifted away from the overlap region according to \Eq{gauge-right-left}.
If the overlap integral defined in \Eq{norm} is small,
the amplitude after normalization of both $\phi^n_L$ on the 
left side of Fig.~\ref{overlap}
and of $\phi^n_R$ at the right side of Fig.~\ref{overlap}
will be very large.  
Perturbation theory will then yield huge corrections to the eigenvalues 
if $\delta L$ has an off diagonal matrix element 
$ < \phi_L |\delta \L | \phi_R >$
connecting regions $X$ and $Y$. 
The pseudospectral instability thus becomes stronger and stronger
as the overlap integral  
$<\phi^m_L|\phi^m_R > $ becomes small compared with the norm of the 
individual eigenvectors. 

Fig.~\ref{deg} illustrates why 
this overlap integral also becomes smaller
as eigenvalues approach degeneracy.
According to the biorthogonality  condition (\ref{norm}), 
if $n \neq m$, the right eigenvector
$\phi^m_R$ should be orthogonal to all the left states $\phi^n_L$. 
Near the degeneracy point, when the two left eigenvectors
almost coincide, the right eigenvectors become nearly 
orthogonal to their {\it own} left counterparts, and the overlap integral 
almost vanishes. Thus, nearly degenerate eigenvalues in a 
non-Hermitian operator 
can cause extreme sensitivity to small off-diagonal 
perturbations, corresponding to large pseudospectral regions
for small $\epsilon$. The same mechanism may have caused 
the unstable spectra
of large non-Hermitian matrices (with random entries connecting sites
arbitrarily far apart) 
observed recently  by Chalker and Mehlig \cite{chalker}.   
 
With this background in mind,
let us consider the pseudospectral 
properties of the convection--diffusion 
operator with random growth
considered in this review.
The pseudospectrum of the convection-diffusion 
operator 
\be
\label{Lcd}
\L_{cd} =  \partial_x^2 + v \partial_x
\ee
has been considered by Reddy and Trefethen \cite{Reddy},
for  ``absorbing''  boundary 
conditions. In our language this situation corresponds to a uniform
growth rate with no population allowed outside of the $[0,d]$ region.
As for an infinite potential well in quantum mechanics, the eigenfunctions
satisfy $\phi(0) = \phi(d) = 0$. 
 For $v= 0$, $\L_{cd}$ is Hermitian with a complete set of orthogonal 
eigenfunctions, indexed by an integer $m$, 
$\phi_m \sim \sin(m \pi x/d)$ with the corresponding 
eigenvalues $\Gamma_m =  - \pi^2 m^2 / d^2$. As $v$ is increased, 
the transformation (\ref{gauge-right-left}) 
gives left and right eigenfunctions
of the form $\phi^{L,R}_m \sim \exp(\pm v x) \sin(m \pi x/d)$, 
leading to peaks near the boundaries, similar 
to those sketched in Fig.~\ref{overlap}.
The spectrum undergoes a rigid shift of  $- v^2/4$ for all $m$,
since the above transformation always 
works  for these boundary conditions. 
The result of this left-right  asymmetry  is, as discussed above, 
a broad 
pseudospectral region. In fact, as shown 
in Ref.~\cite{Reddy} 
(for the case $v=1$), {\it all} points in the complex plane 
contained inside the parabola
\be \label{par}
Re \ z = - (Im \  z)^2
\ee
are contained   
in $\Gamma_\epsilon(\L_{cd})$ for small $\epsilon$,
and the norm in \Eq{def1} diverges
exponentially as 
$d \to \infty$. In the light  of our 
discussion above - this  extreme instability arises from 
small
perturbations of the evolution operator which connect the two bumps at the 
edges, i.e., from small changes of the boundary conditions. 

We show that the results  in this review are in fact unaffected 
by this interesting instability.
First, we note that for {\it periodic}
boundary conditions, the eigenstates  of the non-Hermitian operator 
$\L_{cd}$ (see \Eq{Lcd}) are simply
$\phi_n^{R,L} = \exp(\pm  i k_n  x)$ with $k_n = 2 \pi n/d$ 
and eigenvalues $\Gamma_n = - k_n^2 + i v k_n$,  {\it i.e.},
the eigenvalues lie on the parabolic limit of the pseudospectral 
region illustrated in Eq.~(\ref{par}) for $v=1$.
Since the wave functions  in this case do not have the asymmetric
form implied by Eq.~(\ref{gauge-right-left}), we
expect stability against small perturbations. 

Consider the random matrix model of Sec.~\ref{sec:IV}, now with a change
of a {\it single} $w_j$ to allow boundary conditions similar to those
of Ref.~\cite{Reddy}. Specifically, we insert a factor $\eta$ in front 
of the corner matrix elements of \Eq{matrix-1},
\be
w_N \rightarrow \eta w_N \, .
\ee
By tuning $\eta$ to zero, we break the chain and discard the periodic
boundary conditions. Upon making the same transformation which led to 
\Eq{matrix-Lprime}, we see that all effects of convection vanish
when $\eta =0$, so the eigenvalue spectrum remains real for an 
arbitrary array of convection velocities $g_i$, similar to the results for 
``absorbing'' boundary conditions in the simple convection-diffusion 
problem in Ref.~\cite{Reddy}.
If $\eta$ is set to zero in \Eq{matrix-Lprime}, it is easy to show that a 
'gauge' transformation like (\ref{gauge-right-left}) relates
the eigenfunctions to their $g=0$ values for {\it all} values of $g$,
with a trivial shift of the real $g=0$ eigenvalue spectrum.

The results change dramatically, however, for any $\eta >0$,
no matter how small. Indeed, it is clear from \Eq{matrix-Lprime} 
that the eigenvalue spectrum for nonzero $\eta$ is similar 
to that for $\eta=1$, provided we calculate the spectrum with
effective values of $g^\pm$,
\be
g^{\pm} = g \pm ln(\eta)/N \, ,
\ee
where the $+$ sign refers to the lower left corner matrix element
and the $-$ sign refers to the upper right. All effects of $\eta$
vanish in the continuum limit $N \rightarrow \infty$, and
the eigenvalues approach a universal spectrum like that in 
Fig.~\ref{winding-numbers}, for all $\eta \neq 0$. It is 
straightforward to show, using the methods of Sec.~\ref{sec:IV},
that the eigenfunctions are also unchanged in the large $N$ limit.
We conclude that the results for periodic boundary conditions are
stable to a change in boundary conditions in one dimension,
provided the chain is not actually broken by setting $\eta \equiv 0$.

Our conclusions about the stability of the spectrum and eigenfunctions
for periodic boundary conditions also apply to the simple convection
diffusion operator without randomness. The exceptional sensitivity
to small perturbations of the spectrum for 
``absorbing'' boundary conditions \cite{Reddy} can be understood
in the following way: Since the operators we consider here are
real, all eigenvalues must be real, or occur in complex conjugate
pairs. If, in addition, we impose absorbing boundary conditions, we
exclude all states carrying a real current. 
This exclusion implies real eigenfunctions which insures
that all imaginary parts of the eigenvalues vanish for the lattice
model in the limit $\eta =0$. Replacing this absorbing boundary
condition by a weak link allows the eigenvalues to escape into the
complex plane. The real spectrum studied in \cite{Reddy},
considered as the $\eta \rightarrow 0$ limit of a lattice model, 
involves the {\it simultaneous} coincidence of {\it all} eigenvalues
in pairs (except those at the band edge). This condition of 
extreme defectivity (an $N \times  N$ matrix with only $\sim N/2$ linearly
independent eigenvectors !) is responsible for the exponential instabilities
studied in \cite{Reddy}.

\vspace{1cm}
\bigskip
Acknowledgements: It is a pleasure to acknowledge helpful conversations
with P.W. Brouwer, A. Kudrolli, T. Neicu, Y. Oreg, and L. N. Trefethan.
This work has been supported primarily by the 
Harvard Materials Research Science and Engineering Center 
through Grant No. DMR94-00396, and by the National Science Foundation 
through Grant No. DMR97-14725.

\appendix
\section{Analytical results for the spectra in the homogeneous system}

We give here the expressions for the spectra of a 
2 dimensional system with a homogeneous
growth rate, computed for a square lattice, a triangular lattice,
and the continuum model.
For a homogeneous growth rate the eigenstates can be assumed to be 
plane waves, with ${\mathbf k}$-vectors in the first Brillouin zone. 
For simplicity we set the growth rate everywhere to zero.
(A nonzero constant growth rate merely amounts to a constant shift 
of the eigenvalues $\Gamma({\mathbf k})$.) 
For the square lattice one obtains 
\be
\label{a1}
\Gamma({\mathbf k}) = 2 w \sum_{{\mathbf \nu} = 1}^2 
[ \cosh({\mathbf{g}}\cdot {\mathbf e}_\nu) (\cos({\mathbf k} \cdot {\mathbf e}_\nu) -1 )
+ i \sinh({\mathbf{g}}\cdot {\mathbf e}_\nu) \sin({\mathbf k} \cdot 
{\mathbf e}_\nu)]\, .
\ee
For the triangular lattice the eigenvalues are 
\be
\label{a2}
\Gamma({\mathbf k}) = {4 w\over 3} \sum_{{\mathbf \nu} = 1}^3 
[ \cosh({\mathbf{g}}\cdot {\mathbf e}_\nu) (\cos({\mathbf k} \cdot {\mathbf e}_\nu) -1 )
+ i \sinh({\mathbf{g}}\cdot {\mathbf e}_\nu) \sin({\mathbf k} \cdot 
{\mathbf e}_\nu)]\, ,
\ee
while for the continuum model the expression is simply
\be
\Gamma({\mathbf k})= - D k^2 + i {\mathbf k} \cdot {\mathbf v} \, ,
\ee
a form which agrees with the expansions of \Eq{a1} and \Eq{a2}
for small wave vectors $\mathbf k$.


\begin{references}
\bibitem{ref-frame}
For a moving mask,
as in figure~\ref{setup}, the Fisher equation reads
$\partial c({\mathbf x}',t)/\partial t =
D (\nabla')^2 c({\mathbf x}', t) +
U({\mathbf x}'-{\mathbf v} t) c({\mathbf x}', t)
- b c^2({\mathbf x}', t)$, where ${\mathbf x}'$ is the 
coordinate in the laboratory frame where the growth 
medium is fixed. Upon transforming 
to new spatial coordinates ${\mathbf x} = {\mathbf x'} - {\mathbf v} t$,
in which the mask remains fixed, and then making the 
replacement $c({\mathbf x} + {\mathbf v} t, t) 
\rightarrow c({\mathbf x}, t)$, we obtain 
equation~(\ref{equation-of-motion}).
\bibitem{normdef} The norm of a matrix $A$  is defined here   as $||A|| \equiv
 \sqrt{max. \  eigenvalue \  of  \ A^{\dagger} A}$
\end{references}
\end{document}